\documentclass[review,number,sort&compress]{elsarticle}
\usepackage{amsmath}
\usepackage{amssymb}
\usepackage{siunitx}
\usepackage{tabularx}
\usepackage{adjustbox}
\usepackage{multirow}
\usepackage{rotating}
\usepackage{graphics}
\usepackage{graphicx}
\usepackage{adjustbox}
\usepackage{arydshln}
\usepackage{array}
\usepackage{subfigure}
\usepackage[font=footnotesize,labelfont=bf,singlelinecheck=false]{caption}
\usepackage{lineno}
\usepackage{braket}
\usepackage{float}
\usepackage{hyperref}
\hypersetup{
     colorlinks   = true,
     citecolor    = red
}
\DeclareSIUnit\torr{Torr}
\journal{}%Journal of Nuclear Instruments and Methods: A}
\begin{document}
\begin{frontmatter}
\title{Demonstration of ThGEM-Multiwire Hybrid Charge Readout for Directional Dark Matter Searches} 
\author[]{A. C. Ezeribe\corref{cor1}}%
\ead{a.ezeribe@sheffield.ac.uk} %
\author[]{~C. Eldridge}%
\author[]{~W. Lynch}%
\author[]{~R. R. Marcelo Gregorio}%
\author[]{~A. Scarff}%
\author[]{~and~N. J. C. Spooner}%
\cortext[cor1]{Corresponding author}%
\address{Department of Physics and Astronomy, University of Sheffield, Hounsfield Road S3 7RH, United Kingdom.}

\begin{abstract}
Sensitivities of current directional dark matter search detectors using gas time projection chambers are now constrained by target mass. A ton-scale gas TPC detector will require large charge readout areas. We present a first demonstration of a novel ThGEM-Multiwire hybrid charge readout technology which combines the robust nature and high gas gain of Thick Gaseous Electron Multipliers with lower capacitive noise of a one-plane multiwire charge readout in SF$_6$ target gas.  Measurements performed with this hybrid detector show an ion drift velocity of $138~\pm~10$~\si{\meter\per\second} in a reduced drift field $\text{E/N}$ of 93~$\times~10^{-17}$~\si{\volt\centi\meter\squared} with an effective gas gain of $2470\pm160$ in 20~\si{\torr} of pure SF$_\text{6}$ target gas.
\end{abstract}

\end{frontmatter}

\section{Introduction}\label{IntroSec}
Detection and characterisation of dark matter (DM) - thought to be Weakly Interacting Massive Particles (WIMPs) \cite{bertone2005, cosmicVision2017, partdatagroup} in a direction sensitive nuclear recoil detector with a suitable target material, is a major goal of the DM search community \cite{Ahlen2010,mayet2016, spergel1988, drift2015}. This technology offers the potential to discriminate WIMP candidate events with galactic signature from terrestrial backgrounds/artefacts and hence, can probe below the so-called neutrino floor \cite{Grothaus2014, Ohare2015, Billard2012}. The use of low pressure gas Time Projection Chamber (TPC) technology, in which ionisation electrons from the nuclear recoil tracks are drifted to a charge readout plane and recorded for reconstruction, offers a route to achieving this goal. This is with potentials for low energy threshold and low background operations, including active electron recoil discrimination in the low WIMP mass parameter space. 

The CYGNUS consortium is extensively exploring the feasibility of this technology for a large-scale experiment with aim to search for WIMPs beyond the so-called neutrino floor \cite{ Ahlen2010,CYGNUSPaper}. This builds on previous R$\&$D and DM search results by multiple directional efforts, including DRIFT \cite{driftLowThresh} , NEWAGE \cite{newage2015}, MIMAC \cite{mimac2013}, D$^3$ \cite{d32015}, DM-TPC \cite{dmtpc2014} and CYGNO \cite{cygno2019} collaborations. A feature of interest in DRIFT, for instance, is the use of CS$_2$ gas for primary ionization charge transport through negative ions (NI)  drift, rather than drifting electrons for minimal and thermal scale diffusion \cite{drift2015, ifft2000}. Primary ionization electrons from interactions in the TPC attach rapidly to the electronegative CS$_\text{2}$ to form anions. These anions are drifted towards the readout plane where they are field ionized by the inhomogeneous high electric field  in this region - thereby inducing signal amplification by electron avalanche \cite{ifft2000}. The use of NI drift substantially reduces blurring of the tracks by diffusion \cite{ifft2000, drift2016, drift2017}, and hence saves cost by allowing the possibility for longer drift distances relative to the conventional electron drift concepts. 

Recently, it has been discovered that SF$_6$ \cite{Aleksandrov1988, Grimsrud1985}, which has lower toxicity with improved handling over CS$_2$ \cite{Ezeribe2017}, can also serve as a negative ion TPC gas. This is with a further advantage of formation of a minority charge carrier specie SF$_\text{5}^-$, in addition to the main SF$_\text{6}^-$ charge carrier specie \cite{Phan2017}. Measurement of the arrival time difference between these charge species at the readout plane, allows for identification of the absolute perpendicular distance between an event interaction vertex and the charge readout plane. This characteristic is vital for full rejection of background events emanating from the surfaces of the detector materials. Such event fiducialisation power has been demonstrated using a controlled admixture of O$_2$ gas in a CS$_2$:CF$_4$ based target gas \cite{drift2015, Ifft2014}. 

The higher $^\text{19}$F content in SF$_6$ (relative to CF$_4$) offers a further advantage for improved WIMP-nucleon spin-dependent sensitivity \cite{Tovey2000, Bednyakov2004}.  Studies indicate that stronger avalanche fields are required near the readout planes to achieve field ionization of SF$_6$ anions for electron avalanche due to the higher electron affinity of SF$_\text{6}$ relative to the CS$_\text{2}$ gas \cite{Phan2017}. These strong avalanche fields are outside the operational range of more fragile electrode configurations in the conventional multiwire proportional counter (MWPC) geometry as used in DRIFT \cite{Charpak1979, Ferbel1991, Alner2005}. However, Thick Gaseous Electron Multipliers (ThGEMs) \cite{Sauli2016, Burns2017} have been demonstrated to produce gains of order 10$^3$ in SF$_6$ gas \cite{Phan2017}. Initial results show that a gain of 10$^4$ can be achieved with a triple thin GEM setup \cite{Ishiura2019}. Studies are ongoing to develop more efficient SF$_6$ gas purification \cite{Ezeribe2017} and recycling systems to ensure that minimal or no SF$_6$ gas is released to the atmosphere. This is vital as the greenhouse effect from a given mass of SF$_6$ gas is $\sim$4 orders of magnitude worse than an equal mass of CO$_2$ gas.  

The combination of the high gas gain from ThGEMs and the low capacitance of the multiwires offers a route to achieving a lower operational threshold with potential for a 3-d track reconstruction ability if used with two-plane multiwire configuration. Hence, the possible signal-to-noise ratio that can be achieved in operations with non-hybrid ThGEM or MWPC based TPC technologies with SF$_6$ target gas can be surpassed. 

In this work, we present for the first time, a demonstration of a ThGEM-Multiwire hybrid charge readout technology as a possible candidate for next generation large area, low threshold TPC-based directional dark matter detectors. In this hybrid configuration, field ionization of anions occur on the ThGEM while induced charge signals by the avalanche electrons are read out using wires coupled at a \si{\milli\meter}-scale distance behind the ThGEM. 

\section{Design and Construction of the ThGEM-Multiwire Hybrid Detector}\label{sec:detectorDesign}
\begin{figure}[h!] 
\centering
\subfigure[Detector configuration.]{%
\includegraphics[clip, trim=1.8cm 7.5cm 1.7cm 11.8cm, width=.7\textwidth,height=0.3\textheight]{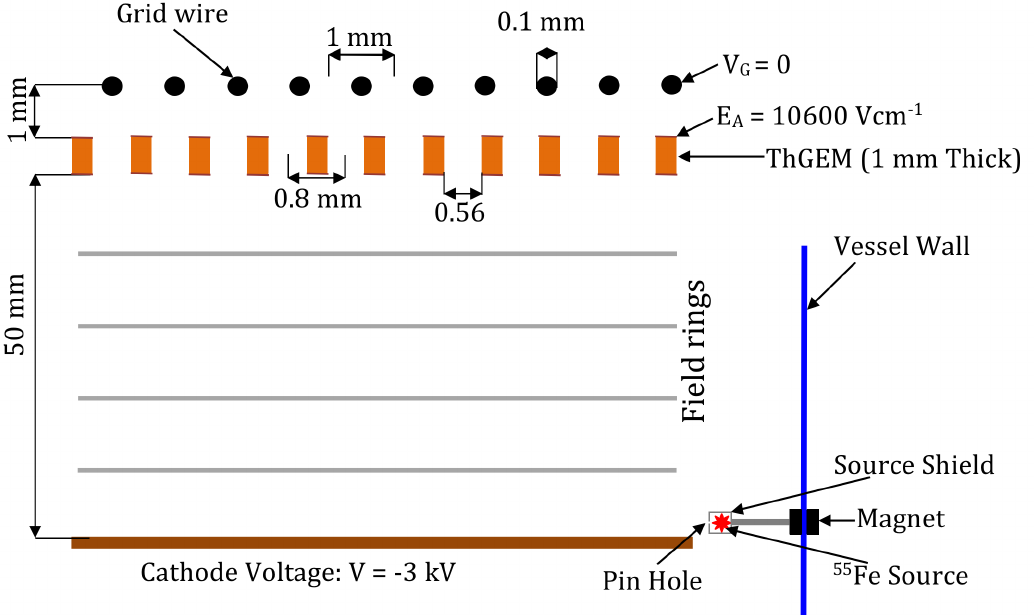}%.5,0.25
\label{fig:detectorConfiguration}
}\hfil
\subfigure[Picture of the detector.]{%
\includegraphics[clip, trim=1.8cm 7.2cm 4.1cm 12.6cm,width=0.9\linewidth,height=0.35\textheight]{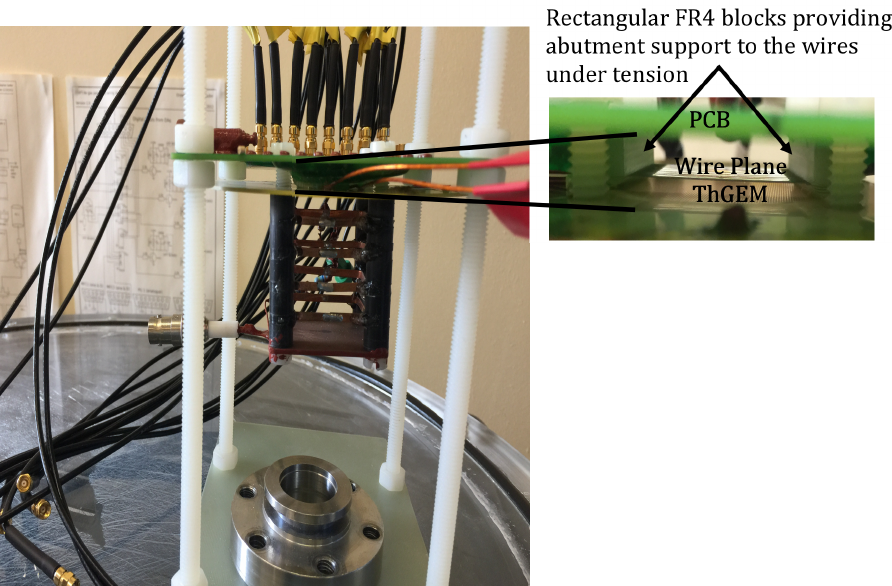}%0.43,0.28
\label{fig:detectorPics}%
}
\caption{Illustration and picture of the hybrid detector. The black dots in \protect\subref{fig:detectorConfiguration} are the 100~\si{\micro\meter} stainless steel wires of the Multiwire-readout while orange rectangles mark the boundaries of the ThGEM holes. Light grey lines in \protect\subref{fig:detectorConfiguration}  are the field rings and the red star is $^{55}$Fe or $^{241}$Am source. The E$_{\text{A}}$ and V$_{\text{G}}$ labels are the avalanche electric field and grid voltage, respectively. In  \protect\subref{fig:detectorPics} is a photograph of the detector when placed on the transparent lid of the vacuum vessel, showing the copper-plate cathode, field cage, ThGEM, wire-plane and SMC cables for connecting the wires to amplifiers (colour online).}
\label{fig:hybrid:illustration}
\end{figure}

The ThGEM-Multiwire hybrid detector technology combines the robust nature and high gas gain of ThGEM readouts with low capacitive noise and the ability to achieve better event track granularity from multiwires.
The hybrid detector used in this work was made from a circular, 1~\si{\milli\meter} thick GEM (sourced from CERN) of  5~\si{\centi\meter} fiducial diameter coupled to a 2~\si{\centi\meter}~$\times$~2~\si{\centi\meter}, one-plane multiwire readout \cite{Ezeribe2018, Ezeribe2016}. Hence, this setup allows for track reconstruction in 2-d by combining charge drift times with $x$-axis track information from the one-plane multiwires. An illustration of the detector configuration with typical operational voltages and a picture of the detector is shown in Figure \ref{fig:hybrid:illustration}. Studies to use two-plane multiwires for full 3-d track reconstruction is topic of future work.

The diameter and pitch of the hexagonally arranged circular ThGEM holes was 0.56~\si{\milli\meter} and 0.8~\si{\milli\meter}, respectively. A cross section of the ThGEM is shown in Figure \ref{fig:detectorConfiguration}. Either side of the ThGEM holes were enclosed by an additional  0.04~\si{\milli\meter} rim, etched on the copper-clads to prevent electrostatic hole-edge discharges and ensure that electric field lines are centred on the ThGEM holes for optimal ion collection and field ionization for the avalanche process. The rim size can affect the performance of a ThGEM based detector. For instance, increasing the rim size of the ThGEM from 0.04~\si{\milli\meter} to 0.09~\si{\milli\meter} in a Garfield simulation \cite{GARFIELDpp}, resulted in 86\si{\percent} loss of initial electrons into the copper cladding and dielectric FR$\text{4}$ material. It is important to point out that this is one of the early set of ThGEMs produced by CERN, so it does not represent an optimal design. The one-plane Multiwire readout was made using 100~\si{\micro\meter} diameter stainless steel wires, placed on a custom made printed circuit board at 1~\si{\milli\meter} pitch. The wire plane was then mounted on the induction side of the ThGEM at a ThGEM-Multiwire separation of 1~\si{\milli\meter}.  The sensitivity of such hybrid detector configuration can be improved in future designs by ensuring that the wire pitch is equal to the ThGEM hole pitch.

A field cage was designed and constructed to maintain a uniform drift field \cite{Blum2008, Knoll2010} within the 2~\si{\centi\meter}~$\times$~2~\si{\centi\meter}~$\times$~5~\si{\centi\meter} detector volume as shown in Figure \ref{fig:detectorPics}. This was achieved by stepping down the high-voltage applied to the copper-plate cathode through a series of five, 33~\si{\mega\ohm} resistors connected to four series of copper field rings. To complete the circuit, the last field cage ring was connected to ground through the fifth successive resistor. The detector was then built by mounting the ThGEM-Multiwire readout on the field-cage with the ThGEM (charge transfer) side of the readout facing the drift volume as shown in Figure \ref{fig:hybrid:illustration}. 

\begin{figure}[h!] 
\centering
\includegraphics[width=1\textwidth,height=0.2\textheight]{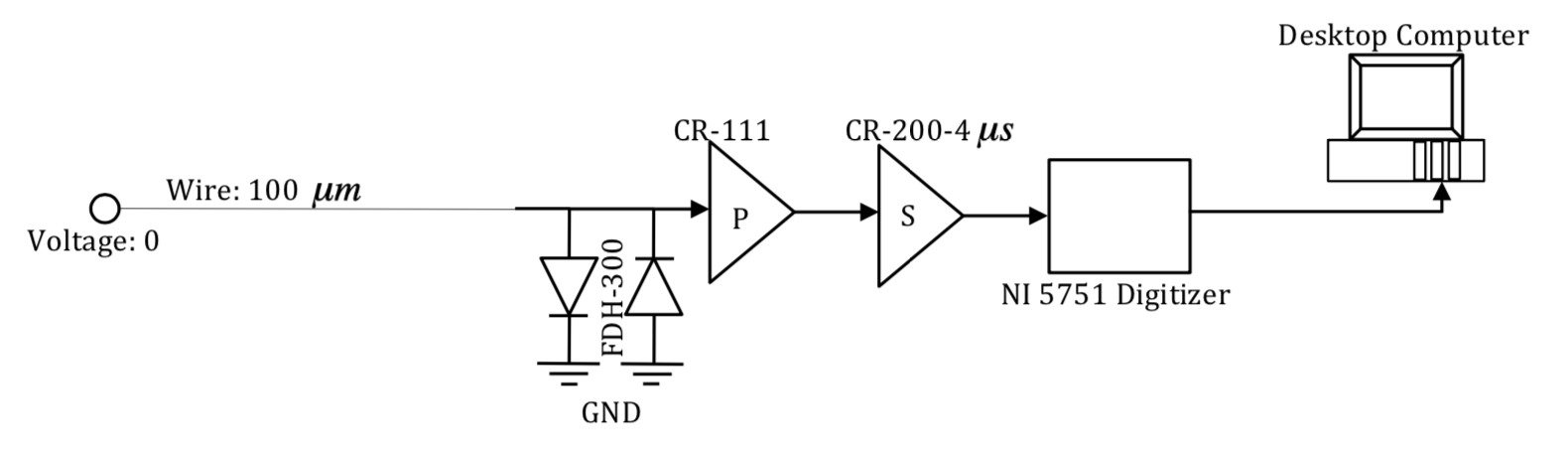}
\caption{Path of the data flow. Arrows show the direction of the data flow.}
\label{fig:dataFlow}
\end{figure}

\begin{figure}[b!] 
\centering
\subfigure[Electric field contour.]{%
\includegraphics[clip, trim=.5cm 0.cm 0.cm 0.1cm, width=0.5\textwidth,height=0.23\textheight]{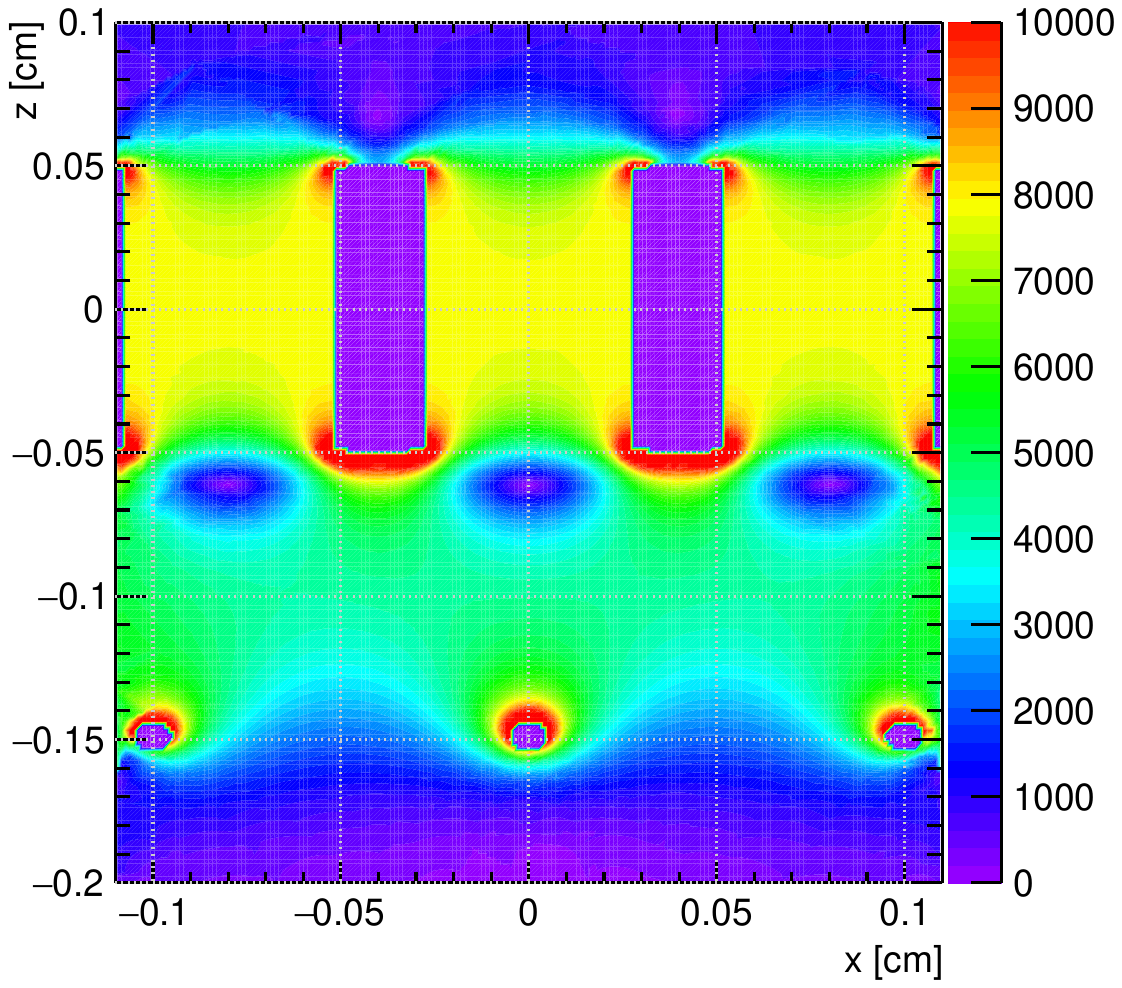}
\label{fig:ThGEMWireContour}
}\hfil
\subfigure[Charge trajectory.]{%
\includegraphics[clip, trim=1.0cm 1.cm 0.1cm 0.2cm, width=0.47\linewidth,height=0.23\textheight]{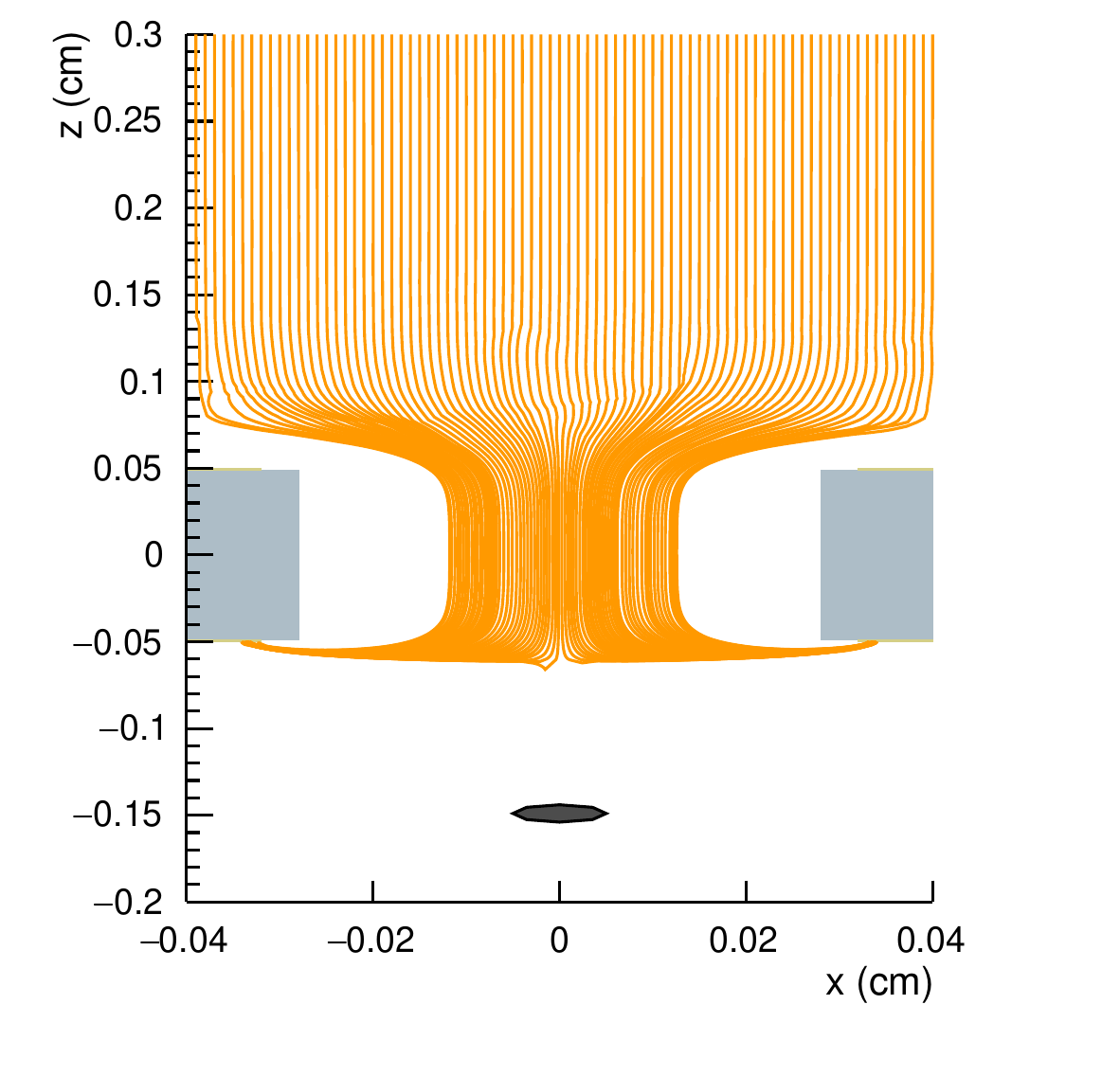}
\label{fig:fieldLines}%
}
\caption{Electric field contour around three ThGEM holes and wires is shown in  \protect\subref{fig:ThGEMWireContour} while the charge trajectory through a ThGEM hole and around a read-out wire is shown in \protect\subref{fig:fieldLines} as obtained with a Garfield simulation by biasing the drift field, charge transfer and induction sides of the ThGEM and read-out wires to $-700$~\si{\volt\per\centi\meter}, $-450$~\si{\volt}, $+600$~\si{\volt}, and 0~\si{\volt}, respectively. The colour scale in \protect\subref{fig:ThGEMWireContour} shows the strength of the electric field measured in \si{\volt\per\centi\meter}.}
\label{fig:GarfieldSim}
\end{figure}

The full data flow path from the read-out wire to storage disk is shown in Figure \ref{fig:dataFlow}. 
During an operation, the charge transfer side of the ThGEM was biased to ensure that the drift field is maintained before the drifting anions are field-ionized for electron avalanche. The avalanche process and signal multiplication was induced by setting the opposite (induction) side of the ThGEM to a sufficient and more positive potential.  By biasing the wire potential to 0~\si{\volt}, avalanche electrons induce equivalent current \cite{Blum2008,Knoll2010,Shockley1938, Ramo1939} on the wires as they follow the charge trajectories to the induction side of the ThGEM as shown in Figure \ref{fig:GarfieldSim}. 
For a simple case where an electrode is set to voltage $V$ while other surrounding electrodes are grounded, the current $i_t$ induced by a charge $e$ moving along a trajectory $x_t$ towards the electrode can be defined as \cite{Blum2008}:
\begin{equation}
i_t =  -\frac{e}{V}E_n[x_t]v_t \,,
\label{eq:i}
\end{equation}  
\nolinebreak
 where $E_n$ is electric field of the electrode when the charge $e$ is removed while $v_t$ is the instantaneous velocity of the charge. The induced charge $Q$ over a given time $t$ can then be determined using:
 \begin{equation}
Q = \int_{0}^{t} i_t dt \,.
\label{eq:q}
\end{equation}  
\nolinebreak
Induced charge signals were used as avalanche electrons can reattach to SF$_{\text{6}}$ to form anions while drifting from the induction side of the ThGEM to the charge collection wires.  To do this, gas gain obtained from charge signals collected on the induction side of the ThGEM were measured and compared to the effective gain from induced charge signals on the wires. This was found to be consistent at similar detector conditions. 

\begin{figure}[h!] 
\centering
\includegraphics[width=0.75\textwidth,height=0.33\textheight]{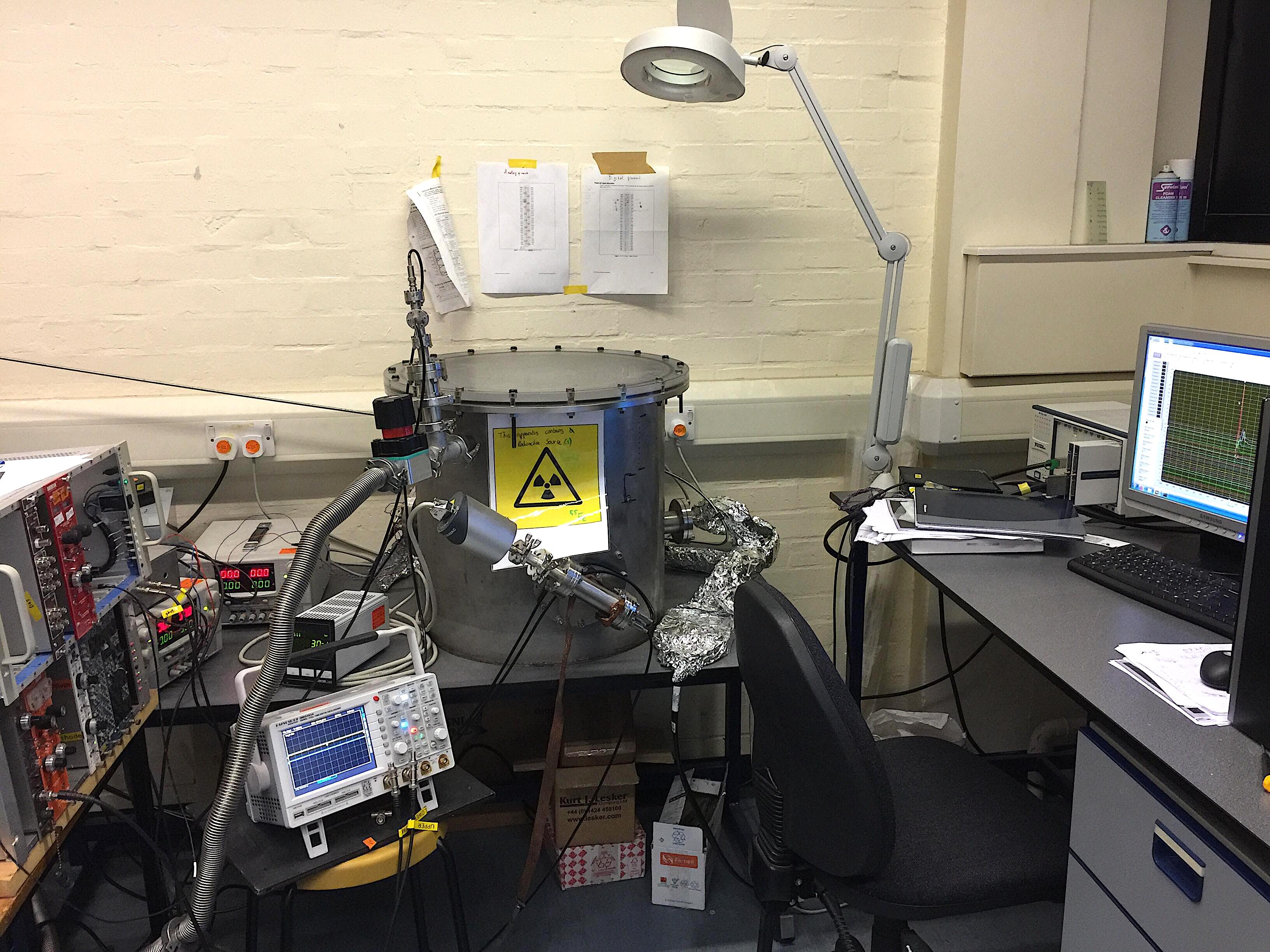}
\caption{Picture of the experimental setup stand showing high voltage power supplies in NIM crates used to bias the copper-plate cathode, ThGEM and wires. The d.c. power supplies that power the amplifiers are also shown with the vacuum vessel, pressure gauge/monitor, NI digitizer and the data computer.}
\label{fig:LabSetup}
\end{figure}

Garfield simulations were used to determine the operational voltage configurations for the detector. For instance, typical operational drift field range of $-300~\si{\volt\per\centi\meter}~\text{to}~-700$~\si{\volt\per\centi\meter}, charge transfer ThGEM voltage of $-450$~\si{\volt} and induction  ThGEM voltage of $+600$~\si{\volt} and wire bias voltage of 0~\si{\volt} were used.  An electric field contour and the expected charge trajectory in this voltage configuration are shown in Figure \ref{fig:GarfieldSim}. The high electric field gradient around the vicinity of the charge transfer side of the ThGEM (see Figure \ref{fig:ThGEMWireContour}) induces the field ionization process. Positive (inverted) charge signals were induced on the surrounding wires by electromagnetic distortions caused by the motion of avalanche electrons towards the wires as they proceed to the induction side of the ThGEM.  These induced currents on the wires were successively amplified using Cremat CR-111 pre-amplifiers and CR-200-4\si{\micro\second} gaussian shaping amplifiers. As shown in Figure \ref{fig:dataFlow}, a pair of grounded inverse parallel FDH-300 diodes were added between the pre-amplifiers and the wires to protect the amplifiers from power surge.  Amplified signals were digitised using an NI 5751 digitizer controlled through a NI PXI-7953R FlexRIO FPGA and saved to disk for analyses. Due to the capacity of the digitizer, only 16 wire channels were instrumented. An FPGA \cite{Farooq2012} and LabVIEW \cite{Elliott2007} based data acquisition system (DAQ) was developed for online data quality monitoring and run control. A picture of the experimental setup stand is shown in Figure \ref{fig:LabSetup}.
The white dual channeled ISEG NHQ 238L NIM cassette high voltage power supply shown in Figure \ref{fig:LabSetup} powered the copper-plate cathode and the charge transfer side of the ThGEM while the red dual channeled Bertan 377P power supply biased the induction side of the ThGEM and the read-out wires. The BST PSM 2/2A d.c. power supplies shown in Figure \ref{fig:LabSetup} were used to bias the Cremat amplifiers which are located inside the vacuum vessel to minimise noise distortions. A Leybold Ceravac CTR-101 pressure gauge was used to monitor the pressure of the 96~\si{\litre} vacuum vessel.

\section{Detector Calibration}\label{sec:detectorPerformance}
Gain measurement was performed to investigate the detector performance using X-rays from the electron capture decay of $^\text{55}$Fe source to $^\text{55}$Mn. As shown in Figure \ref{fig:detectorConfiguration} the source was positioned close to the copper-plate cathode to irradiate the detector fiducial volume. To achieve this, the source was bonded to the tip of a 5~\si{\centi\meter} M6 nylon studding glued to a Neodymium disc magnet and attached to the inside wall of the vessel. This was magnetically coupled to a second magnet on the outside vessel wall which was used to control the source position in the vessel. Ionization electrons from X-ray interactions with the target gas through photoelectric effect - attach to the electronegative SF$_\text{6}$ to form anions. As described in Sections \ref{IntroSec} and \ref{sec:detectorDesign}, these anions drift in a uniform field to the ThGEM for field ionization and electron avalanche. Signal pulses induced on the wires from this process were amplified and recorded to disk at a frequency of 1~\si{\mega\hertz} per channel, without any hardware trigger as the pulses were small (for instance, $>$5~\si{\milli\volt} in 30~\si{\torr} of SF$_{\text{6}}$).  All signal pulses on each of the 16 wires with amplitude $>$5~\si{\milli\volt} threshold were analysed in the $^\text{55}$Fe runs to reject pedestal and electronic noise.  Pulses with $>$3~\si{\volt} amplitude were not included in the analysis to remove sparks and events that saturate the amplifiers and the digitizer. 

\begin{figure}[h!]
\centering
\subfigure[Data spectrum from the Fe55 run.]{%
\includegraphics[clip, trim=.2cm 0.2cm 0.2cm 0.2cm, width=0.49\linewidth,height=0.28\textheight]{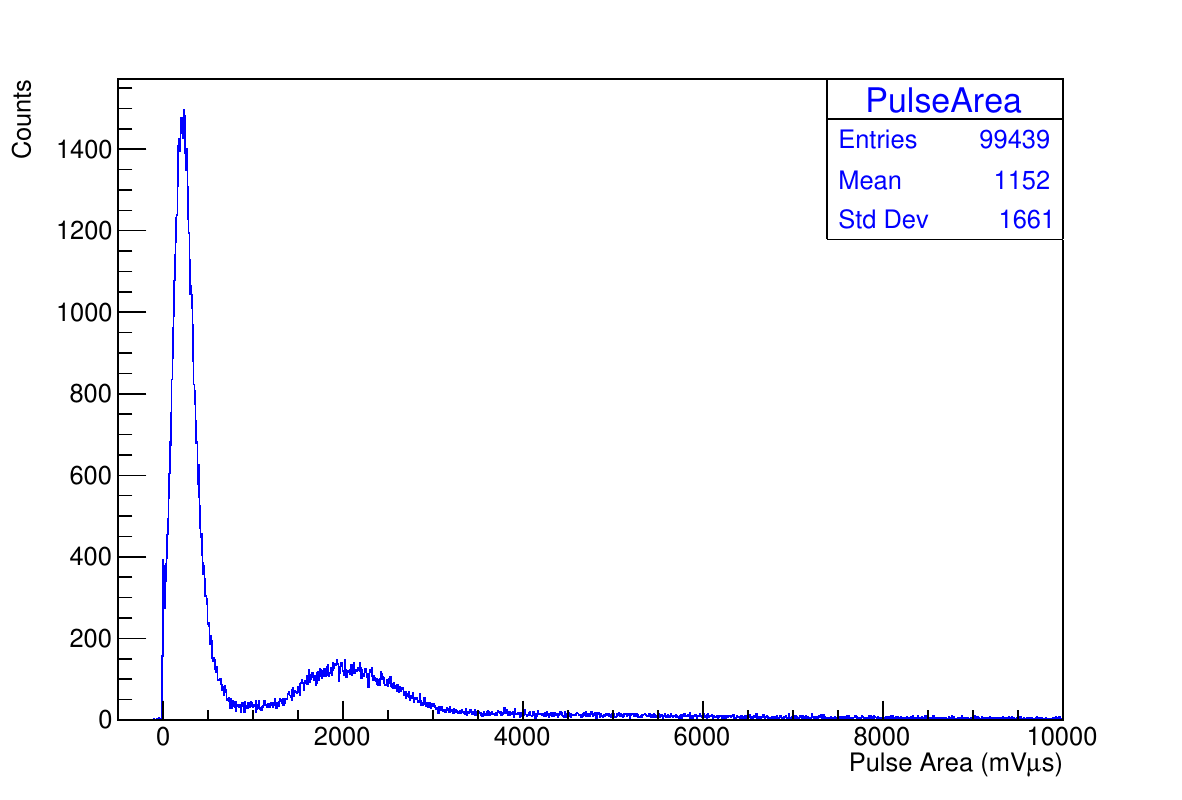}
\label{fig:rawfe55}%
}\hfil
\subfigure[Zoomed-in version of the Fe55 peak.]{%
\includegraphics[clip, trim=.2cm 0.2cm 0.2cm 0.2cm, width=0.49\linewidth,height=0.28\textheight]{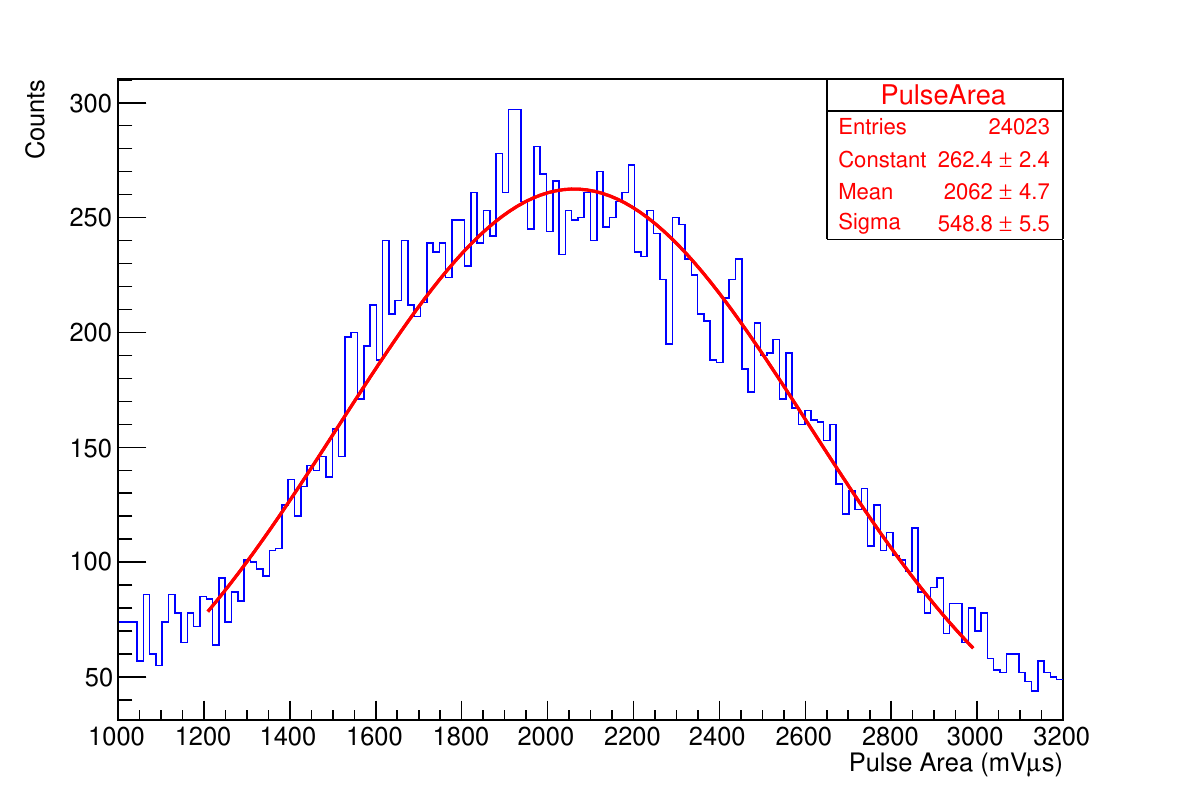}
\label{fig:Fe55Fit}%
}
\caption{Energy spectrum of 5.9~\si{\kilo\electronvolt} X-rays from $^\text{55}$Fe source. All the events that passed the analyses cuts are shown in \protect\subref{fig:rawfe55}. This includes electronic noise and the  X-ray events from the $^\text{55}$Fe source while \protect\subref{fig:Fe55Fit} zooms-in on peak of the X-ray events. Red line is a gaussian fit to the X-ray events (colour online).}
\label{fig:Fe55}
\end{figure}

\begin{figure}[h!]  
\centering
\includegraphics[clip, trim=.8cm 0.5cm 0.8cm 0.8cm, width=0.58\textwidth,height=0.35\textheight]{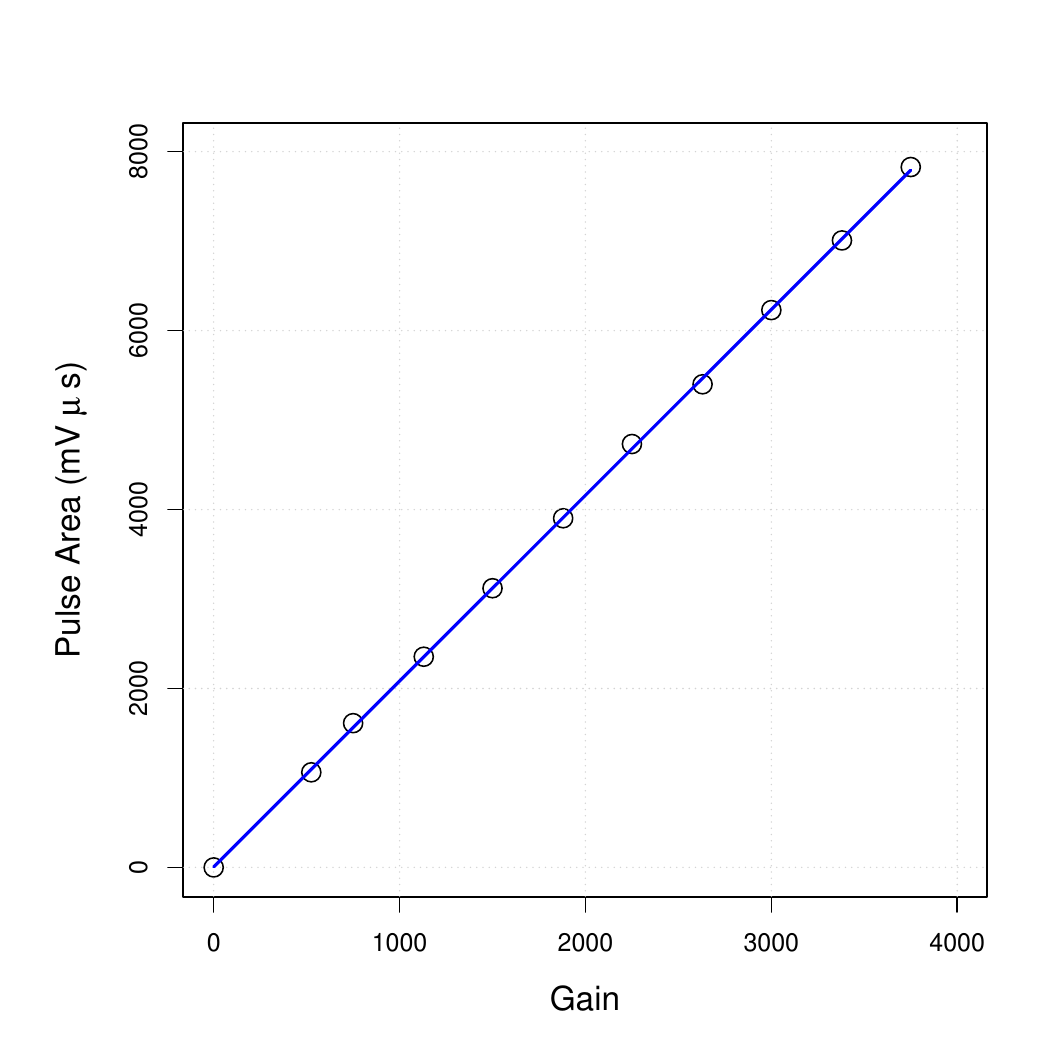}
\caption{Integral charge from amplifier calibration pulses shown as a function of expected gas gain from $^{\text{55}}$Fe X-ray interactions. The blue solid line is a linear fit on the data (colour online).}
\label{fig:gainCal}
\end{figure}

Background events due to radioactive decays from detector materials (at relevant energy) which could mimic the $<$2 wire channels trigger from the $^\text{55}$Fe X-ray interactions are expected to be minimal in the short exposure time (about 2~\si{\hour}) of this source run. To determine the pulse area, any charge that passed the analysis threshold on each of the signal channels were integrated from  the 10~\si{\micro\second} time before the pulse rising edge crosses the threshold to the 10~\si{\micro\second} time after the pulse falling edge crosses the threshold. The energy spectrum of events that passed the analysis cuts is shown in Figure \ref{fig:Fe55}. 
The peak of a gaussian fit on the observed spectrum of the X-ray data is 2062~$\pm$~5~\si{\milli\volt\micro\second}.  To understand the detector gas gain, the amplifier gain was calibrated using test pulses of 14~\si{\milli\volt} (minimum output voltage of the pulser) and 20~\si{\milli\volt} to 90~\si{\milli\volt} amplitude at 10~\si{\milli\volt} interval. To do this, each of the test pulse signals was connected to the test input of the pre-amplifiers. The \si{\milli\volt}-scale test pulses were converted to charge signals through a 1~\si{\pico\farad} test capacitor of the pre-amplifiers. The charge output of the pre-amplifier was then coupled to the shaping amplifier for further amplification and shaping. The shaped pulse signals were then digitized, saved to disk and analysed using the same analyses algorithm used in the X-ray data shown in Figure \ref{fig:Fe55}.  As in the X-ray data, a gaussian was fitted on each of the amplifier calibration data and the peak of the fits were extracted and analysed as a function of the expected detector gas gain from 5.9~\si{\kilo\electronvolt} X-rays from $^{\text{55}}$Fe exposures. Results from the calibration pulse analyses are shown as a function of the expected detector gas gain in Figure \ref{fig:gainCal}.
The expected gas gain shown in Figure \ref{fig:gainCal} was determined by converting each of the observed test charge to their equivalent number of ion pairs (NIPs) using: 
\begin{equation}
NIPs_o =  \frac{CV}{e} \,,
\label{eq:nipso}
\end{equation}  
\nolinebreak
where $NIPs_o$ is the observed number of ion pairs, $C$ is the test capacitance of the amplifier, $V$ is the test pulse in \si{\milli\volt} and $e$ is the electronic charge. The expected number of ion pairs ($NIPs_e$) from an $^{\text{55}}$Fe X-ray was determined using:
\begin{equation}
NIPs_e =  \frac{E}{W} \,,
\label{eq:nipse}
\end{equation}  
\nolinebreak
where $E$ is the 5.9~\si{\kilo\electronvolt} energy of $^\text{55}$Fe X-rays while $W$ is the mean energy required to create an electron-ion pair in SF$_\text{6}$ target gas - measured to be 35.45~\si{\electronvolt} in Ref. \cite{Hilal1987}. This implies that the $^\text{55}$Fe X-ray will produce 166.4 electron-ion pairs after an interaction with the target gas before the electron avalanche. Hence, the detector gas gain is defined as the ratio of the $NIPs_o$ to the $NIPs_e$. Using parameters of the linear fit ($\text{Q}_\text{A} = 2G + 6$) in Figure \ref{fig:gainCal}, where $\text{G}$ and $\text{Q}_\text{A}$ are the effective gas gain and the integral charge, respectively. These parameters and the observed integral charge from the $^\text{55}$Fe analyses yield a gain of 1028~$\pm$~3. 
\begin{table}[h!] 
\normalsize
\centering
\caption{Detector parameters used in the detector calibration run performed in 30~\si{\torr} of pure SF$_{\text{6}}$ target gas. The \text{V}, \text{E},  \text{$\Delta\text{V}_\text{A}$} and $\text{E}_\text{A}$ parameters are cathode voltage, drift field, avalanche voltage and avalanche fields, respectively. The \text{E/N} ($\text{E}_\text{A}/\text{N}$) is the reduced drift (avalanche) field while $\text{N}$ is the gas density computed to be $9.7\times10^{17}$ \si{\per\cubic\centi\meter}.} 
\begin{adjustbox}{width= 12.0 cm}
\renewcommand{\arraystretch}{.8}
\begin{tabular}{l*{6}{c}r}
\hline \hline %\\
\text{V (\si{\volt})} &  \text{E} (\si{\kilo\volt\per\centi\meter}) 	& $\text{E/N}$ ($10^{-17}$ \si{\volt\centi\meter\squared})  	&  	\text{$\Delta\text{V}_\text{A}$} (\si{\volt}) 	& $\text{E}_\text{A}$ (\si{\kilo\volt\per\centi\meter})  & $\text{E}_\text{A}\text{/N}$ ($10^{-17}$ \si{\volt\centi\meter\squared}) \\ %\\
\hline 
3000	&        0.6     &     		62		&  1055 &	   10.6	        &    	    1092     	    \\ %\\	 
\hline \hline
\end{tabular} 
\end{adjustbox}
\label{table:Fe55Calibration}
\vspace*{0.15 cm}
\end{table} 
For details of drift and avalanche fields used in this detector calibration run, see Table \ref{table:Fe55Calibration}. This is using a drift (avalanche) field of 600~\si{\volt\per\centi\meter} (10.6~\si{\kilo\volt\per\centi\meter}) in 30~\si{\torr} of pure SF$_\text{6}$ gas - equivalent to $\text{E/N}$ and $\text{E}_A/\text{N}$ of 62~$\times~10^{-17}$~\si{\volt\centi\meter\squared} and 1092~$\times~10^{-17}$~\si{\volt\centi\meter\squared}, respectively. Here, $\text{E/N}$ is the reduced drift field, $\text{E}_\text{A}/\text{N}$ is the reduced avalanche field and N is the gas density computed to be $9.7\times10^{17}$ \si{\per\cubic\centi\meter}, for more on the N parameter, see section \ref{sec:mobility}. Hence, the calibration gain result does not represent the highest achievable gain of the detector as higher reduced avalanche fields with sufficient reduced drift field configurations can yield higher effective gas gains. The investigation of the highest achievable effective gain of the detector is beyond the scope this paper.

\section{Ionization Tracking Capabilities}\label{sec:tracking}
To further understand the detector performance and demonstrate the ionization tracking capabilities of the detector, the $^\text{55}$Fe source discussed in Section \ref{sec:detectorPerformance} was replaced with an $^\text{241}$Am source which emits 5.5~\si{\mega\electronvolt} alphas as it decays to $^\text{237}$Np. This is such that the mean free path of alphas from the source was few \si{\milli\meter} away from the cathode. As discussed in Sections \ref{sec:detectorDesign} and \ref{sec:detectorPerformance}, ionization electrons along the interaction track of the ionizing alpha, as it traverses the detector fiducial volume attach to the electronegative SF$_\text{6}$ to form anions. These anions were drifted in the uniform electric field to the high field region of the detector readout for field ionization and subsequent electron avalanche.  Signal pulses induced on the wires by this process were pre-amplified, shaped, digitized and saved to disk for analyses. 

\subsection{Events Pre-Selection Analysis and Data Quality}\label{sec:dataquality}
Seven alpha source runs were performed at different reduced drift fields to validate the dependency of the field ionization process on the gradient of the ThGEM charge transfer field relative to the drift field. This is using a constant reduced avalanche field of 1449~$\pm$~1~$\times~10^{-17}$~\si{\volt\centi\meter\squared} as shown in Table \ref{table:ionmobility}, except in the seventh $\text{E}_\text{A}/\text{N}$ field run where it was increased by $\sim$1\% to 1461~$\pm$~1~$\times~10^{-17}$~\si{\volt\centi\meter\squared} to investigate the detector response to a higher $\text{E}_\text{A}/\text{N}$ field. The expectation is that this small increase in the reduced avalanche field should increase the effective gain of the detector.
\begin{table}[b!] 
\normalsize
\centering
\caption{Detector parameters used in the  ionization tracking studies performed in 20~\si{\torr} of pure SF$_{\text{6}}$ target gas. The \text{V}, \text{E},  \text{$\Delta\text{V}_\text{A}$} and $\text{E}_\text{A}$ parameters are cathode voltage, drift field, avalanche voltage and avalanche fields, respectively. The \text{E/N} ($\text{E}_\text{A}/\text{N}$) is the reduced drift (avalanche) field while $\text{N}$ is the gas density computed to be $6.5\times10^{17}$ \si{\per\cubic\centi\meter}.} 
\begin{adjustbox}{width= 12.0 cm}
\renewcommand{\arraystretch}{.7}
%\tiny
\begin{tabular}{l*{6}{c}r}
\hline \hline %\\
\text{V (\si{\volt})} &  \text{E} (\si{\kilo\volt\per\centi\meter}) 	& $\text{E/N}$ ($10^{-17}$ \si{\volt\centi\meter\squared})  	&  	\text{$\Delta\text{V}_\text{A}$} (\si{\volt}) 	& $\text{E}_\text{A}$ (\si{\kilo\volt\per\centi\meter})  & $\text{E}_\text{A}\text{/N}$ ($10^{-17}$ \si{\volt\centi\meter\squared}) \\ %\\
\hline 
1800		& 	0.36      &     	56	&	933		&	   9.33	  &    	    1449     	    \\ \\	 

2000		&	0.40      &    	62	&	933		&	   9.33 	&   	    1449     	    \\ \\   

2200		&	0.44      &     	68	&	933		&	   9.33 	&    	    1449     	    \\ \\	 

2400		&	0.48      &    	75	&	933		&	   9.33 	&   	    1449     	    \\ \\ 

2600		&	0.52      &     	81	&	933		&	   9.33 	&    	    1449     	    \\ \\	 

2800		&	0.56      &    	87	&	933		&	   9.33 	&   	    1449     	    \\ \\
	
3000		&	0.60      &    	93	&	941		&	   9.41	&   	    1461     	    \\ %\\	 
\hline \hline
\end{tabular} 
\end{adjustbox}
\label{table:ionmobility}
\vspace*{0.1 cm}
\end{table} 

\begin{figure}[h!] 
\centering
\includegraphics[width=1\linewidth,height=0.4\textheight]{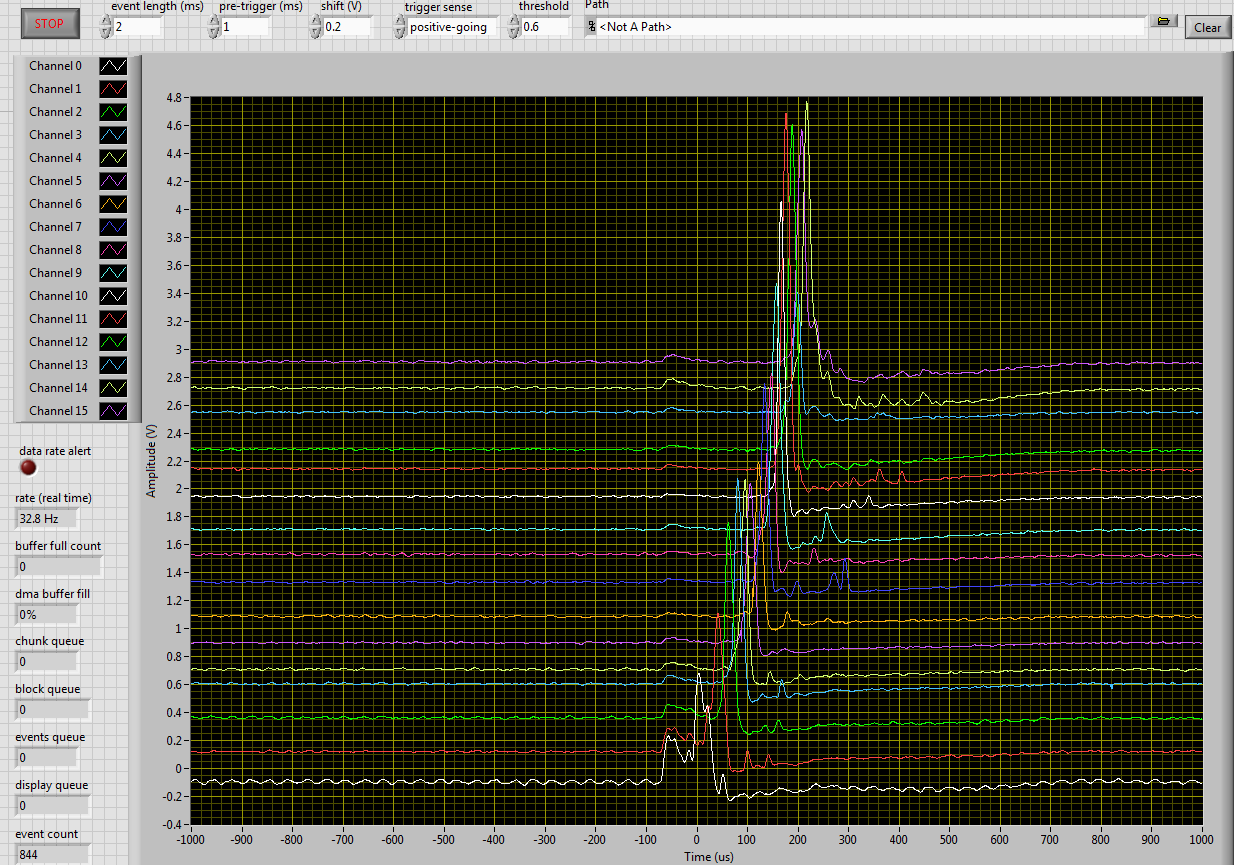}
\caption{An example from part of 5.5~\si{\mega\electronvolt} alpha track observed in the hybrid detector with 30~\si{\torr} of pure SF$_{\text{6}}$ target gas. The full alpha track is expected to be 27.9~\si{\centi\meter} long on average, as determined from SRIM \cite{Ziegler2010} simulations. Channel~0 is closer to the $^\text{241}$Am source.}
\label{fig:alphatracks}
\end{figure}

An example of a raw alpha track oriented towards the cathode as seen on the LabVIEW based DAQ is shown in Figure \ref{fig:alphatracks}.
The alpha track is preceded by a low-energy track, likely from radioactive decays in the detector materials within the opposite vicinity of the source location. For this alpha track, the source was placed  few \si{\milli\meter} away from the ThGEM to observe a clear charge arrival time delay from tracks oriented towards the cathode. It can be seen that the low-energy track is oriented along the expected mean free path of the source-alphas hence, the resulting ionization charge arrived the readout at about same time. The clear observation of delays between induced charge signals on adjacent wires of the alpha track due to the expected slow negative ion drift properties is of more importance to the work reported here. The wire shown as Channel~0 on the DAQ is closer to the alpha source.  The 210~\si{\micro\second} delay between channels~0 and ~15 of the alpha track signal shown in Figure \ref{fig:alphatracks} is due to the drift times of the anions which depends on the incident angle (within the source subtended solid angle) of the alpha track. The observed wiggle on each of the DAQ channel after the main alpha signal pulse is due to amplifier responses and so were not included in the analysis.

\begin{figure}[h!] 
\centering
\subfigure[Pulse amplitude.]{%
\includegraphics[clip, trim=0.1cm 0.6cm 1.2cm 2.2cm,width=0.49\linewidth,height=0.28\textheight]{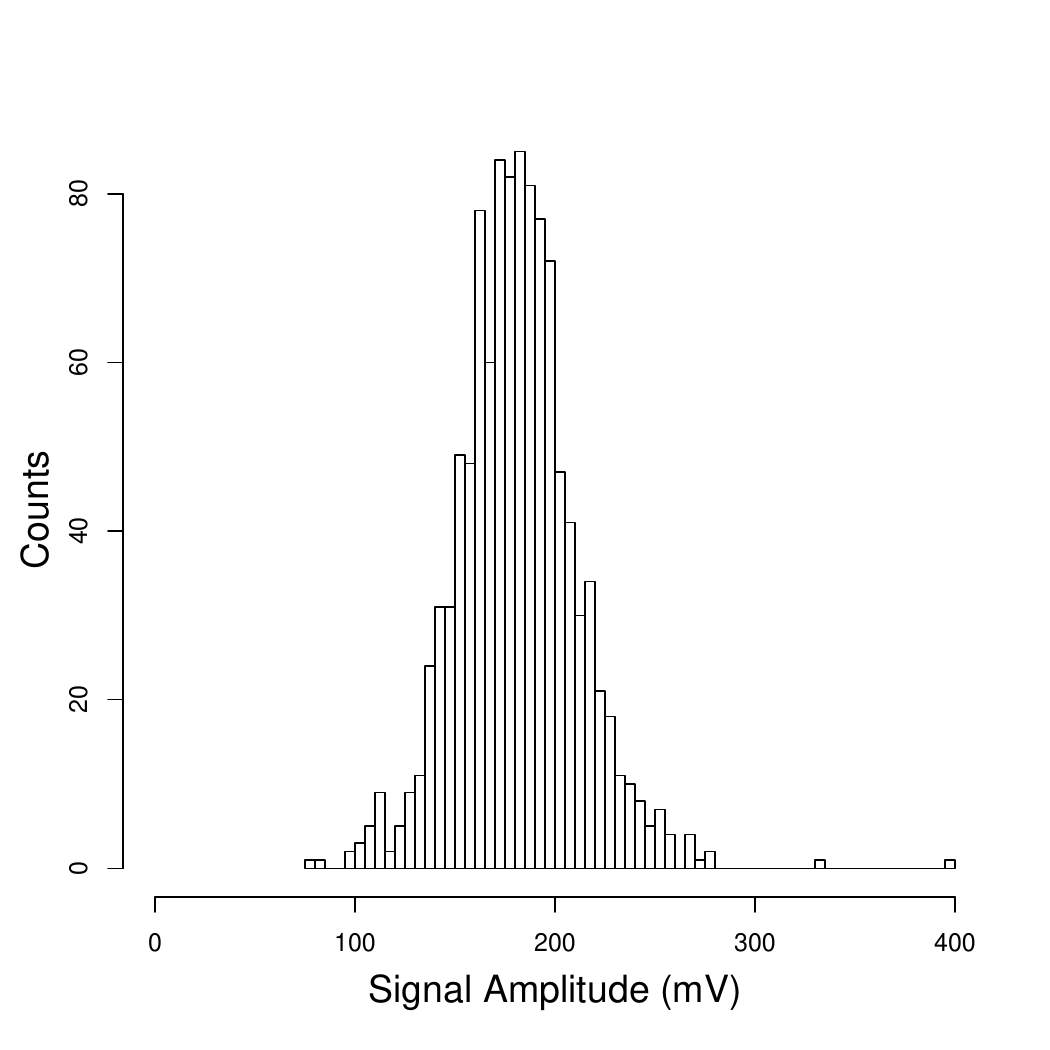}
\label{fig:pulseAmplitude}%
}\hfil
\subfigure[Full width at half maximum.]{%
\includegraphics[clip, trim=0.1cm 0.6cm 1.2cm 2.2cm,width=0.49\linewidth,height=0.28\textheight]{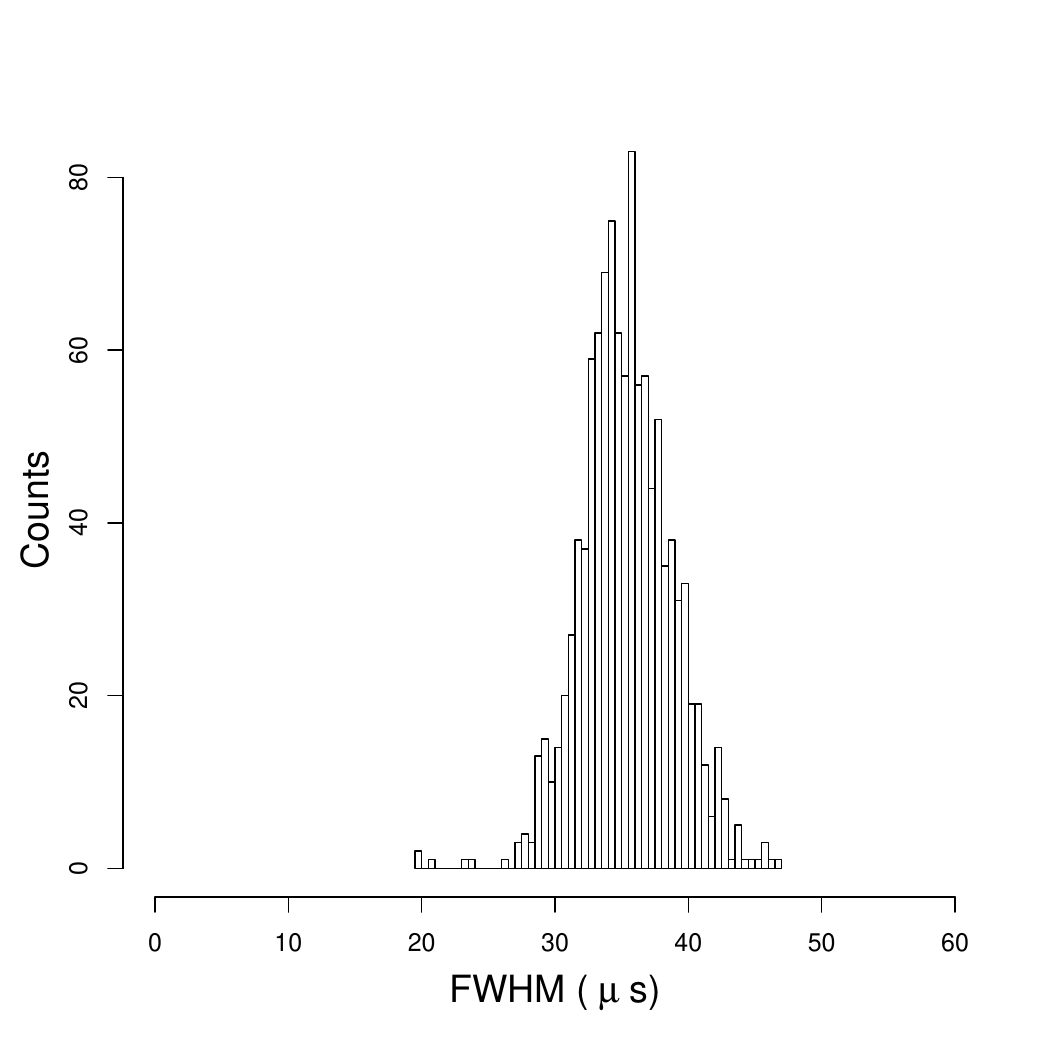}
\label{fig:fwhm}%
}
\caption{Distribution of pulse amplitude and full width at half maximum observed from alpha tracks at a reduced drift field of 56~$\times~10^{-17}$~\si{\volt\centi\meter\squared} and avalanche field of 9.33~\si{\kilo\volt\per\centi\meter} in 20~\si{\torr} of pure SF$_{\text{6}}$ target gas.}
\label{fig:paramDistribution}
\end{figure}

Events with $>$40~\si{\milli\volt} pulse amplitude threshold were analysed further. This is to remove pedestal and electronics noise from the analysis. As discussed in Section \ref{sec:detectorPerformance}, events with $>$3~\si{\volt}  pulse amplitude were also removed. Sparks and other noise pulses with short rise-times were rejected by selecting only events with $>$9~\si{\micro\second} rise-time and full-width at half maximum (FWHM) of 8~\si{\micro\second} to 60~\si{\micro\second}. An example of the pulse amplitude and FWHM obtained from a 56~$\times~10^{-17}$~\si{\volt\centi\meter\squared} reduced drift field run (see Table \ref{table:ionmobility} for more details)  in 20~\si{\torr} of pure SF$_{\text{6}}$ gas are shown in Figure \ref{fig:paramDistribution}.
The average pulse amplitude (FWHM) observed from this run is 181~$\pm$~2~\si{\milli\volt} (36~$\pm$~0.2~\si{\micro\second}). This long FWHM and slow rise-time is consistent with expectations from negative ion drift in the electronegative SF$_\text{6}$ target gas. 

\begin{figure}[h!] 
\centering
\subfigure[Charge at 68~$\times~10^{-17}$~\si{\volt\centi\meter\squared} $\text{E/N}$ mode.]{%
\includegraphics[clip, trim=0.1cm 0.5cm .8cm 2.cm,width=0.49\linewidth,height=0.32\textheight]{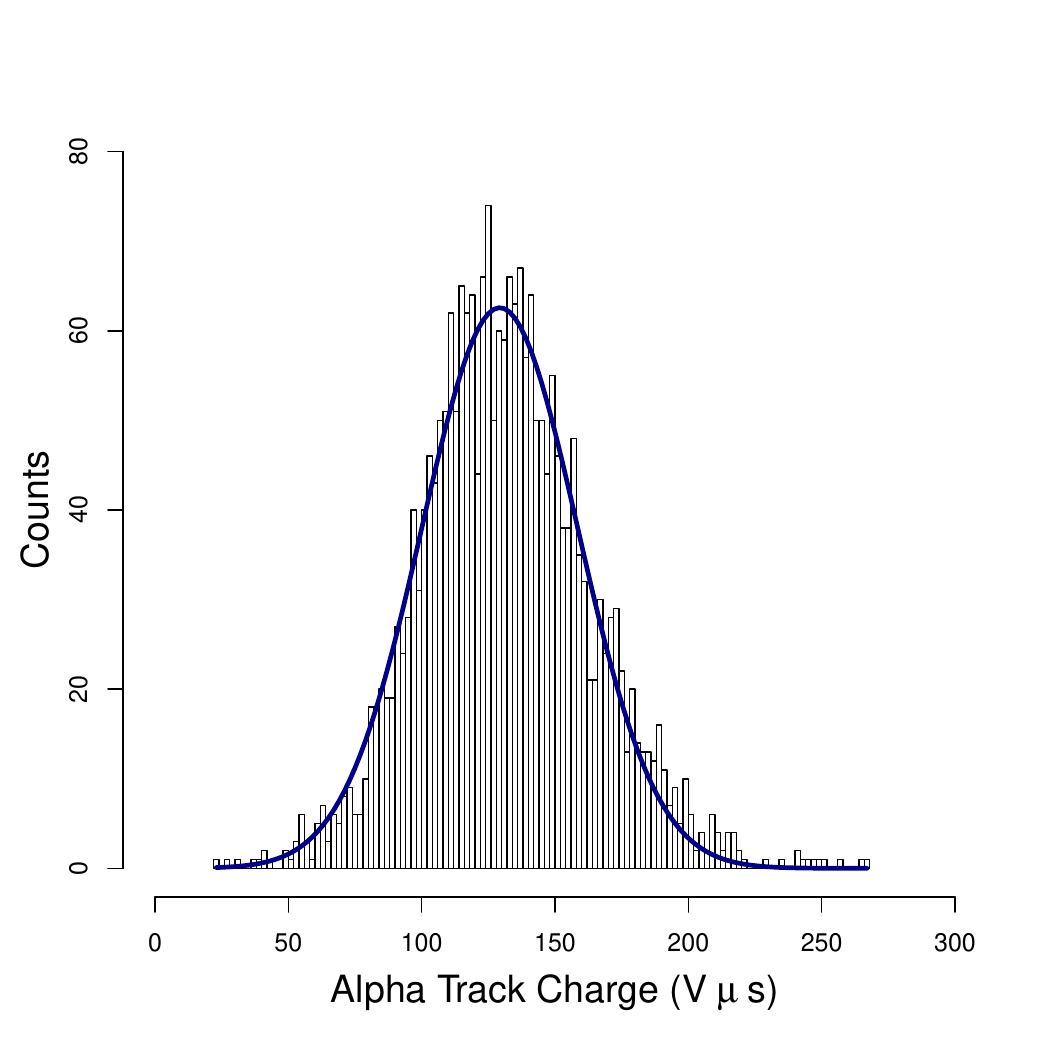}
\label{fig:pulseArea68}%
}\hfil
\subfigure[Charge at 87~$\times~10^{-17}$~\si{\volt\centi\meter\squared} $\text{E/N}$ mode.]{%
\includegraphics[clip, trim=0.1cm 0.5cm 1.cm 2.cm,width=0.49\linewidth,height=0.32\textheight]{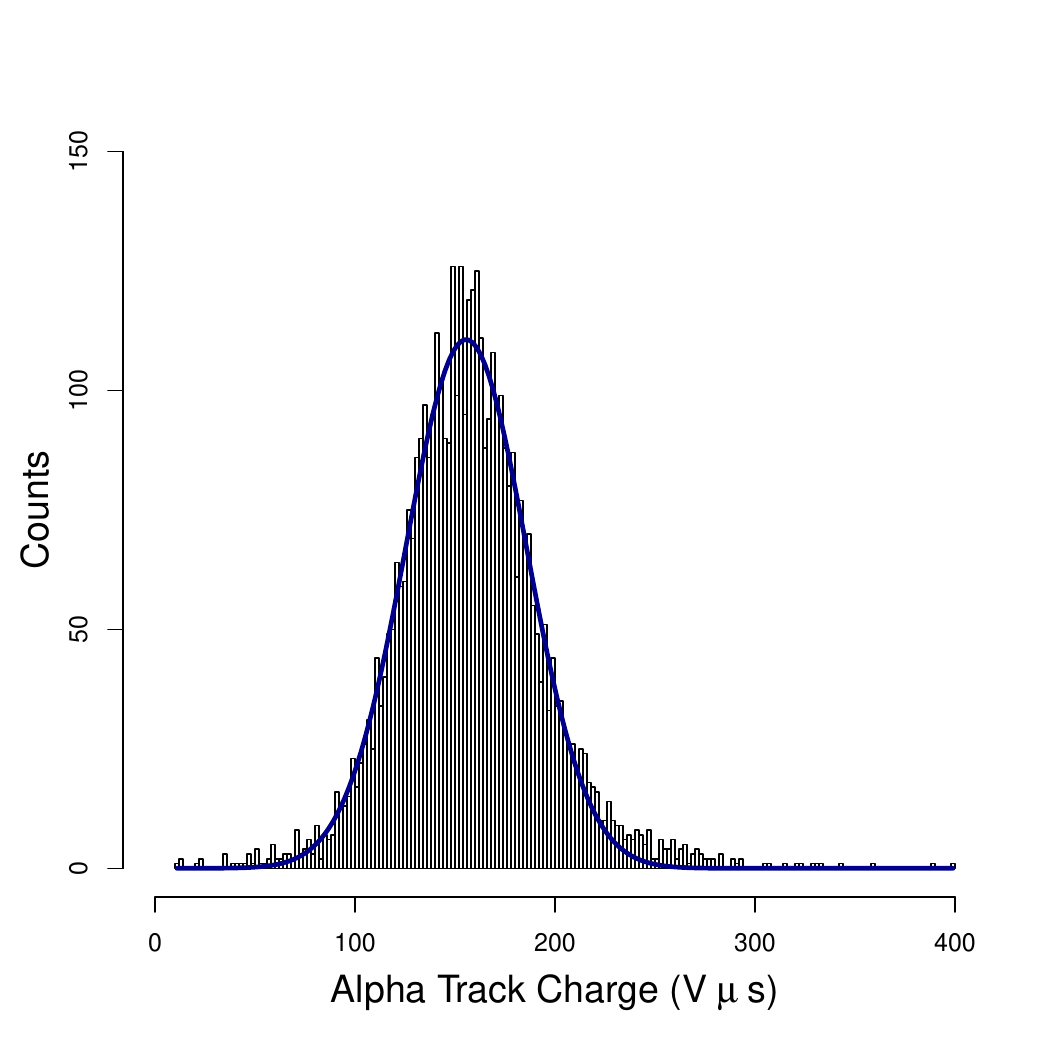}
\label{fig:pulseArea87}%
}
\caption{Distribution of total integral charge from alpha tracks as observed from reduced drift field runs of 68~$\times~10^{-17}$~\si{\volt\centi\meter\squared} and 87~$\times~10^{-17}$~\si{\volt\centi\meter\squared}, respectively. These data were recorded at an avalanche field of 9.33~\si{\kilo\volt\per\centi\meter} in 20~\si{\torr} of pure SF$_{\text{6}}$ target gas. Dark blue lines are gaussian fits on the data (colour online).}
\label{fig:pulseArea}
\end{figure}

As discussed in Section \ref{sec:detectorPerformance}, the effective detector gas gain in each of these runs was measured using alpha events as in Ref. \cite{Burns2017}. To do this, a cumulative integral charge in a fixed time window, around the trigger time for a given signal channel was computed. The maximum from this computation was recorded as the channel charge integral. The sum of this charge integral over all the 16 signal channels is the total integral charge for a given alpha track. This method helps to remove the effect of the pedestal noise from the integral charge computations.  Samples of total integral charge distributions as observed from two of these ionization tracking runs are shown in Figure \ref{fig:pulseArea}. 
There are no visible pedestal noise in Figure \ref{fig:pulseArea} as seen in Figure \ref{fig:rawfe55} due to the lower exposure-times in the alpha runs because the source activity is 2 orders of magnitude higher than that of the X-ray runs in Section \ref{sec:detectorPerformance}. Also, as described above, the charge threshold and cuts applied in the alpha analyses are more stringent for the pedestal noise than in the X-ray data analyses. As described in Section \ref{sec:detectorPerformance},  the peak of the gaussian fits in Figures \ref{fig:pulseArea68} and \ref{fig:pulseArea87} were extracted and used as the effective track integral charge for the effective gain computations. The mean integral charge extracted from the fits in Figure \ref{fig:pulseArea}, are 133~$\pm$~1~\si{\volt\micro\second} and 158~$\pm$~2~\si{\volt\micro\second} for reduced drift field runs of 68~$\times~10^{-17}$~\si{\volt\centi\meter\squared} and 87~$\times~10^{-17}$~\si{\volt\centi\meter\squared}, respectively.  Similar analyses were performed on the remaining 5 alpha runs taking at different reduced drift fields. 

\begin{figure}[b!]  
\centering
\includegraphics[clip, trim=1.0cm 0.5cm 0.5cm 1.5cm, width=0.65\textwidth,height=0.4\textheight]{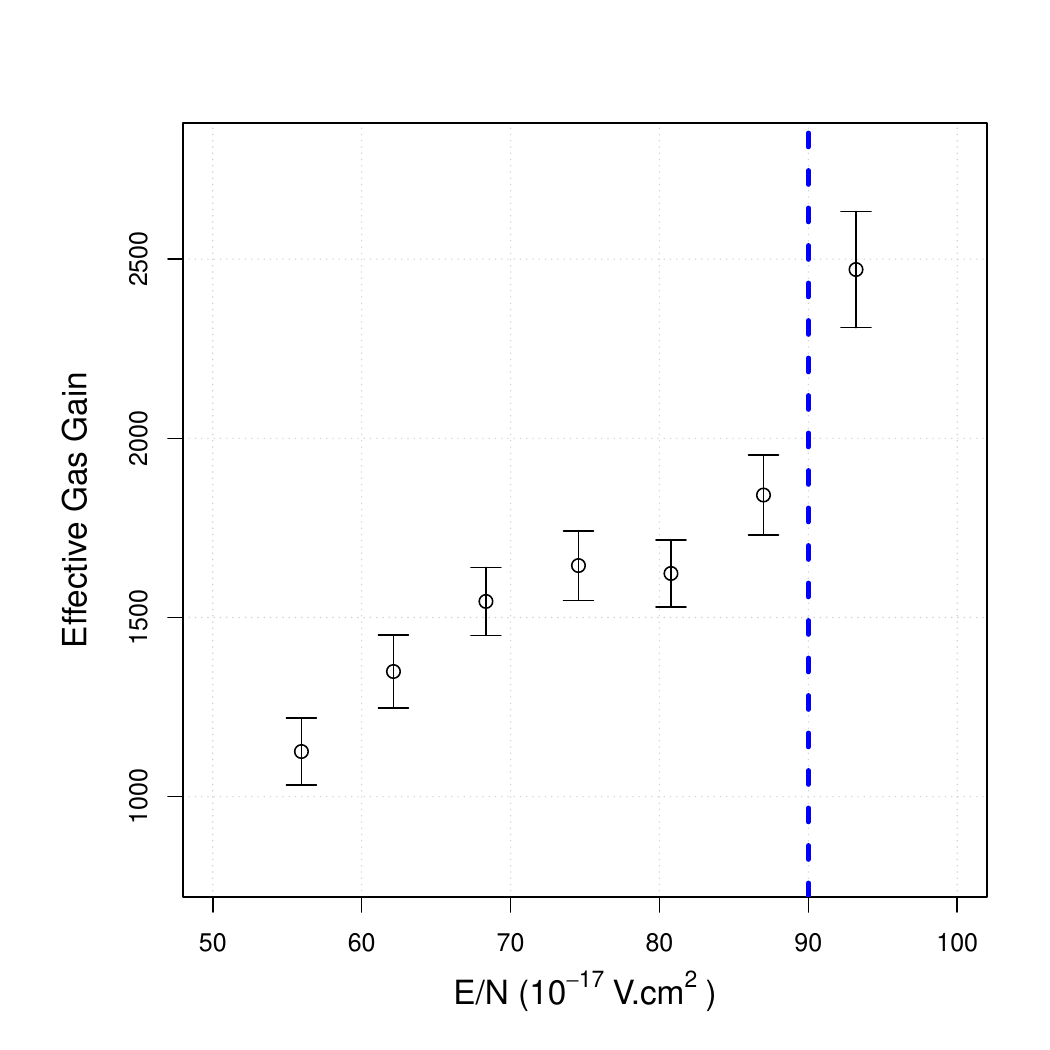}
\caption{Effective gas gains in 20~\si{\torr} of pure SF$_{\text{6}}$ gas shown as a function of reduced drift field. Reduced avalanche fields used in the left and right regions (before and after) of the dashed blue line is 1449 ~$\times~10^{-17}$~\si{\volt\centi\meter\squared} and 1461~$\times~10^{-17}$~\si{\volt\centi\meter\squared}, respectively. Results are from $^{241}$Am alpha tracks.}
\label{fig:alphaGain}
\end{figure}

To convert these results to an effective gain measurement, a SRIM simulation was performed to determine the fraction of the 5.5~\si{\mega\electronvolt} alpha energy that was deposited within the detector fiducial volume as the tracks are expected to be longer than the detector width. This was found to be 0.22~\si{\mega\electronvolt} in average, which translates to 4\% of the total alpha energy. Using this average alpha deposited energy and calibration results from analyses of the pulse calibration runs described in Section \ref{sec:detectorPerformance}, the effective gas gain in each of the alpha runs was determined. The effective gain results from these measurements are shown as a function of the reduced drift fields in Figure \ref{fig:alphaGain}.
The effective gas gain results shown in Figure \ref{fig:alphaGain} increase with the reduced drift field as expected (at low reduced drift fields). This is due to a more positive field gradient in the charge transfer region of the ThGEM as the reduced drift field increases - resulting in better negative ion transparency. The observed effective gas gain in Figure \ref{fig:alphaGain} plateaued from the 75~$\times~10^{-17}$~\si{\volt\centi\meter\squared} to 87~$\times~10^{-17}$~\si{\volt\centi\meter\squared} reduced drift field runs. This is consistent with expectations from reaching the optimal negative ion-drift transparency for the constant reduced avalanche field. The observed rise in the effective gas gain for the reduced drift field run of 93~$\times~10^{-17}$~\si{\volt\centi\meter\squared} (at optimal ion-drift transparency) is due to the higher avalanche field in this run as shown in Table  \ref{table:ionmobility}. However, the drift velocity and mobility of anions are independent of the effective gas gain so these observed plateau and rise in the detector gain should not affect our measurements. These results indicate that a gas gain of 2.5~$\times~10^3$ is feasible with the ThGEM-multiwire hybrid setup in SF$_\text{6}$ target gas.

\subsection{Drift Velocity and Mobility Measurements for SF$_{\text{6}}$ Anions}\label{sec:mobility}
\begin{figure}[b!]  
\centering
\includegraphics[clip, trim=2cm 12.cm 1.3cm 4.8cm, width=0.75\textwidth,height=0.36\textheight]{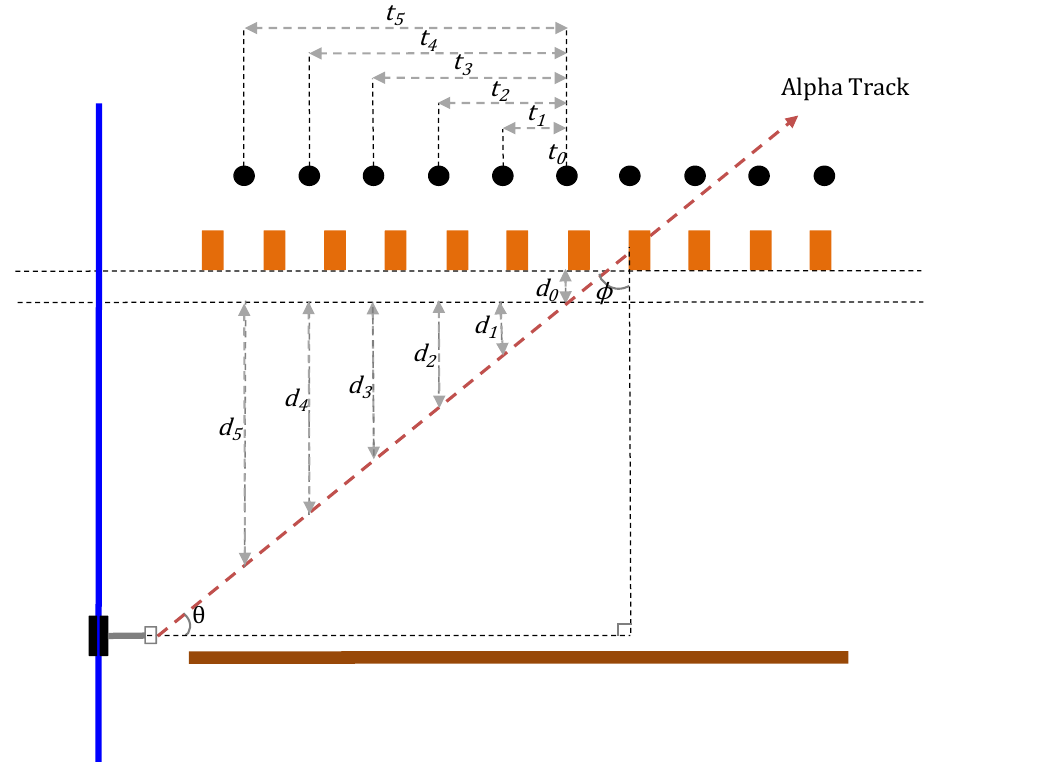}
\caption{Illustration of alpha track selection cuts and parametizations used in the SF$_\text{6}$ anion drift velocity and mobility measurements. Angles $\theta$ and $\phi$ are $81\protect\substack{+1 \\ -0.5}$\si{\degree} and $9\protect\substack{+0.5 \\ -1}$\si{\degree}, respectively.}
\label{fig:mobilityAnalysis}
\end{figure}

To extract the drift velocity and mobility of SF$_\text{6}$ anions, only alpha tracks that made angles of $>81\substack{+1 \\ -0.5}$\si{\degree} with the cathode were selected from each of the runs described in Section \ref{sec:tracking}. For details of drift and avalanche field configurations for each of the runs, see Table \ref{table:ionmobility}. As shown in Figure \ref{fig:mobilityAnalysis}, these tracks were expected to induce signals only on the first 8 wires of the detector. 
Hence, events that recorded hits on these 8 wires, only were selected for further analysis. A hit here is charge induced on a wire that produced a pulse amplitude that passed the analyses threshold. Between 0.2\% to 2.3\% of the total events on disk in the ionization tracking runs passed these cuts and were used in the ion drift velocity and mobility measurements. 

It can be seen in Figure \ref{fig:mobilityAnalysis} that there is no drift distance for the most probable ionization charge that could end on the 7$^{\text{th}}$ and 8$^{\text{th}}$ wires due to the proximity of the interaction vertex to the field ionization region of the ThGEM and readout wire plane, respectively. Hence, signals on these two wires were not included in the drift velocity and mobility computation. The charge arrival time information from the 6$^{\text{th}}$ wire served as the reference for charge drift times recorded on the first 5 wires.

The vertical drift distances (d$_\text{1}$, d$_\text{2}$, d$_\text{3}$, d$_\text{4}$ and d$_\text{5}$) between the expected start positions of charge clusters arriving on the first 5 wires and the start position of charge cluster arriving on the 6$^{\text{th}}$ wire (after traveling a d$_\text{0}$ distance) was determined using $\text{x}_\text{i}\tan\theta$, see Figure \ref{fig:mobilityAnalysis} for more details. Here, $\text{x}_\text{i}$ is the horizontal distance of the first 5 wires relative to the 6$^{\text{th}}$ wire, so $\text{i}$ is 1, 2, ..., 5 while $\theta$ is the track mean free path-cathode angle determined to be $81\protect\substack{+1 \\ -0.5}$\si{\degree}.  Charge drift times (t$_\text{1}$, t$_\text{2}$, t$_\text{3}$, t$_\text{4}$ and t$_\text{5}$) for charge clusters arriving on these first 5 wires were computed as the temporal separation between their respective charge cluster arrival times and the arrival time, t$_\text{0}$, of the charge cluster recorded on the reference 6$^{\text{th}}$ wire. The gradient of a linear fit on a vertical drift distance vs drift time plot from a given run was recorded as the drift velocity, $v_d$, for that run. 

\begin{figure}[h!] 
\centering
\subfigure[Drift velocity. ]{
\includegraphics[clip, trim=0.5cm 0.5cm 0.1cm 2.0cm,width=0.485\linewidth,height=0.32\textheight]{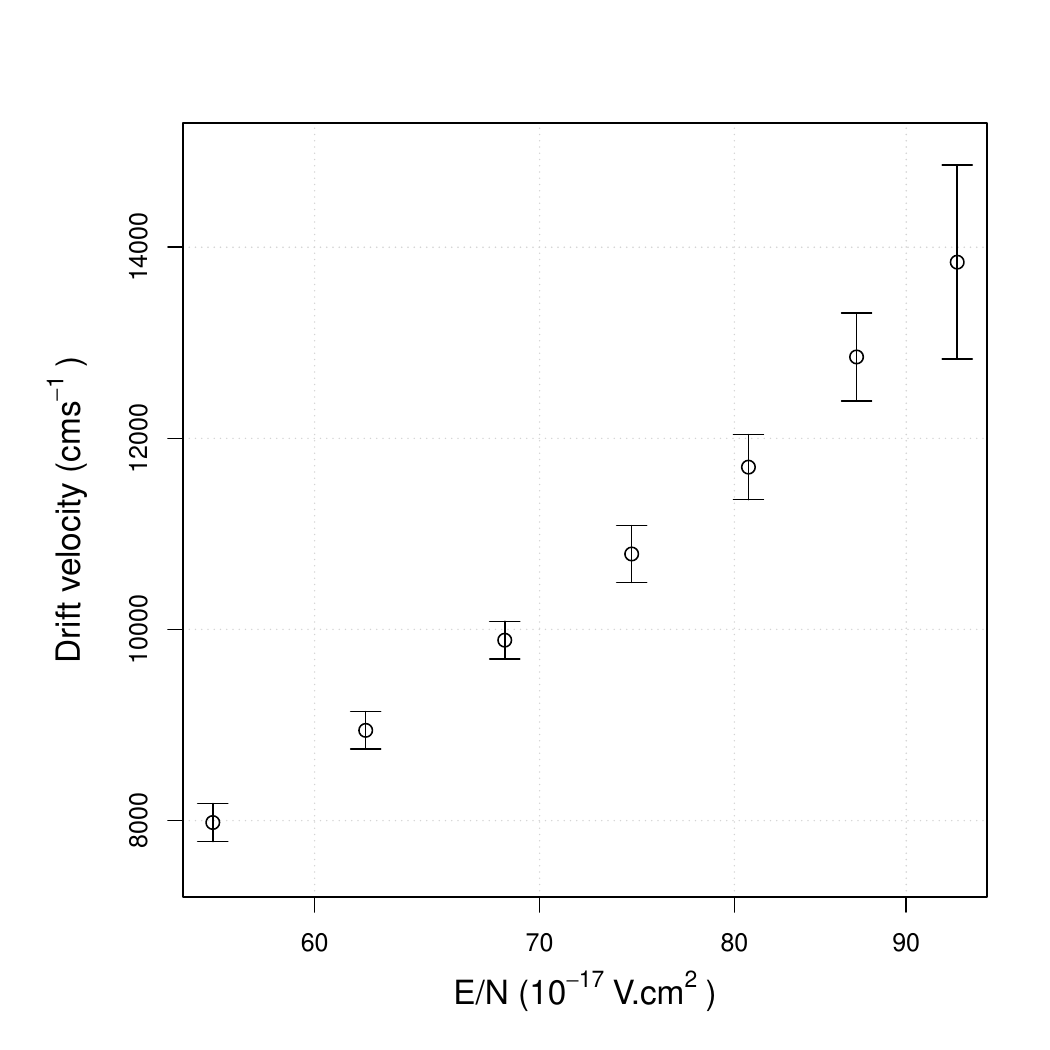}
\label{fig:driftVelocity}%
}\hfil
\subfigure[Reduced mobility.]{%
\includegraphics[clip, trim=0.5cm 0.5cm 0.1cm 2.0cm,width=0.485\linewidth,height=0.32\textheight]{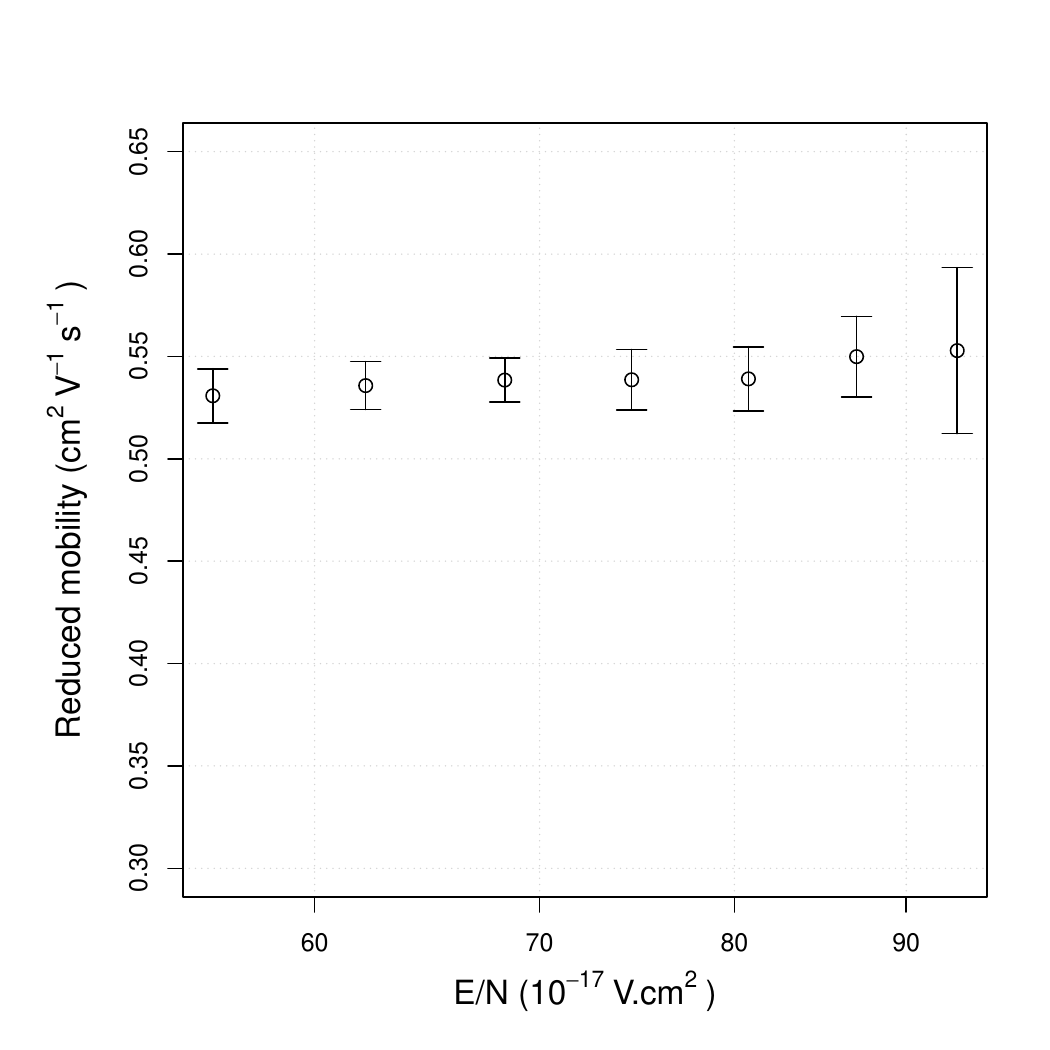}
\label{fig:reducedMobility}%
}
\caption{Drift velocity and reduced mobility for SF$_{\text{6}}$ anions shown as a function of reduced drift field. Only statistical uncertainties are quoted. See Table \protect\ref{table:ionmobility} and \protect\ref{table:ionmobilityresults} for more details on detector parameters used in these measurements and results, respectively.}
\label{fig:resultMobility}
\end{figure}

The mobility, $\mu$, is defined in terms of $v_d$ as:
\begin{equation}
\mu = \frac{v_d }{E} \,,
\label{eq:mobility} 
\end{equation} 
\nolinebreak
where $E$ is the drift field. This can be converted to the reduced mobility $\mu_0$ by considering the operational gas density in each of the measurements for better comparison with other measurements using:
\begin{equation}
\mu_0 = \frac{v_d }{E}\frac{N}{N_0} \,.
\label{eq:reducedMobility} 
\end{equation} 
\nolinebreak
As mentioned in Section \ref{sec:detectorPerformance}, $N$ is the gas density defined as $\frac{\rho A}{Mm}$ in \si{\per\cubic\centi\meter}, where $\rho$ is the gas density for the gas pressure used in a given measurement, $A$ is the Avagadro's constant $6.0221\times10^{23}$~\si{\per\mol} and $Mm$ is the molar mass of SF$_{\text{6}}$ given as 146.06~\si{\gram\per\mol}. The $N_0$ (the Loschmidt constant) is the $N$ parameter computed at the standard temperature (0~\si{\celsius}) and pressure (1 atm or 760~\si{\torr}) which can be evaluated to obtain $2.6868\times10^{19}$ \si{\per\cubic\centi\meter}. Results from the $v_d$ and $\mu_0$ measurements are shown in Table \ref{table:ionmobilityresults} and Figure \ref{fig:resultMobility} as a function of reduced drift fields for a given run in 20~\si{\torr} of SF$_\text{6}$ target gas.
\begin{table}[h!] 
\normalsize
\centering
\caption{Reduced drift field in each of the runs shown with the measured effective gas gain $\text{G}$, drift velocity $v_d$ and reduced mobility $\protect\mu_0$. These measurements were performed in 20~\si{\torr} of pure SF$_{\text{6}}$ target gas.} 
\begin{adjustbox}{width= 12.0 cm}
\setlength{\tabcolsep}{20pt}
\renewcommand{\arraystretch}{.7}
%\tiny
\begin{tabular}{l*{7}{c}r}
%\\
\hline \hline %\\
$\text{E/N}$ ($10^{-17}$\si{\volt\centi\meter\squared})  &  \text{G}  & $v_d$~(\si{\meter\per\second})     &  $\protect\mu_0$~(\si{\centi\meter\squared\per\volt\per\second})  \\ 
\hline 
%\\
56		& 	1130~$\pm$~90		&		80~$\pm$~2 		&	 0.531~$\pm$~0.013	\\ \\	 

62		&	1350	~$\pm$~100		&		89~$\pm$~2  		&	 0.536~$\pm$~0.012	\\ \\   

68		&  	1550~$\pm$~100		&		99~$\pm$~2   		&	0.538~$\pm$~0.011	\\ \\	 

75		&	1650~$\pm$~100		&		108~$\pm$~3  		&	0.539~$\pm$~0.015	\\ \\ 

81		&	1620~$\pm$~90		&		117~$\pm$~3    	&	0.539~$\pm$~0.016	 \\ \\	 
	
87	 	& 	1840	~$\pm$~110		&		129~$\pm$~5   		&	0.550~$\pm$~0.020	\\ \\
	
93		&	2470~$\pm$~160		&		138~$\pm$~10   	&	0.553~$\pm$~0.041	\\
\hline \hline
\end{tabular} 
\end{adjustbox}
\label{table:ionmobilityresults}
\vspace*{0.015 cm}
\end{table} 

It can be seen in Figure \ref{fig:driftVelocity} that the observed drift velocity increases with reduced drift field as expected.  These observed $v_d$ results are consistent within errors with SF$_\text{6}^-$ results in Ref. \cite{Phan2017} for the relevant reduced drift field used in these measurements. The large uncertainties  ($\pm$2~to~$\pm$10~\si{\meter\per\second}) on these results are mainly due to systematics from relative wire-ThGEM hole positions and uncertainties in the estimation of the $\theta$ and $\phi$ angles for the selected alpha tracks.  However, the observed reduced ion mobilities for the respective reduced drift field runs shown in Table \ref{table:ionmobilityresults} and Figure \ref{fig:reducedMobility} are consistent within errors with SF$_\text{6}^-$ results reported in Refs. \cite{Phan2017, Fleming1969, Patterson1970, Urquijo1991}.

\section{Conclusion}
A ThGEM-Multiwire based hybrid time projection chamber (TPC) detector was designed, built and tested for the first time. Effective gas gain measured from the performance tests of the hybrid detector are in the range of $1130~\pm~90$ to $2470\pm160$ at a reduced drift field $\text{E/N}$ range of 56~$\times~10^{-17}$~\si{\volt\centi\meter\squared} to 93~$\times~10^{-17}$~\si{\volt\centi\meter\squared} in 20~\si{\torr} of pure SF$_{\text{6}}$ target gas. Using the hybrid detector, the drift velocity and reduced ion mobility of SF$_\text{6}$ anions were measured at this reduced drift field range. The observed drift velocity (reduced mobility) results were found to be between $80~\pm~2$~\si{\meter\per\second} and $138~\pm~10$~\si{\meter\per\second} ($0.53~\pm~0.01$~\si{\centi\meter\squared\per\volt\per\second} and $0.55~\pm~0.04$~\si{\centi\meter\squared\per\volt\per\second}) in this reduced drift field range. The drift velocity and reduced ion mobility results from these measurements are consistent within errors with other published measurements in Refs. \cite{Phan2017,Fleming1969,Patterson1970,Urquijo1991}. 

Hence, the ThGEM-Multiwire technology has the potential to serve as a robust, low noise charge readout with known fine grain track resolution in the future massive directional dark  matter detector known as CYGNUS-TPC.
 
 \section*{Acknowledgements}
 We would like to thank K. Miuchi, D. Loomba and S. Vahsen for their helpful comments. This work was supported by the Science and Technology Facilities Council through ST/P00573X/1 and ST/K001337/1 grants.

\end{document}